\documentclass[aps,prb,notitlepage]{revtex4-1}
\usepackage{graphicx}
\usepackage{graphics}
\usepackage{amssymb,amsmath}

\begin{document}
\title{Elasticity effects on the stability of growing films.}
\author{Daniel Walgraef}
\email[E-mail me at:]{ dwalgraef@ifisc.uib-csic.es}
\affiliation{
IFISC (CSIC-UIB), Instituto de F\'isica Interdisciplinar y Sistemas Complejos, E-07122 Palma de Mallorca, Spain (http://www.ifisc.uib.es).}

\begin{abstract}
It is shown how the combination of atomic deposition and nonlinear diffusion may lead, below a critical temperature, to the growth of nonuniform layers on a substrate. The dynamics of such a system is of the Cahn-Hilliard type, supplemented by reaction terms representing adsorption-desorption processes. The instability of uniform layers leads to the formation of nanostructures which correspond to regular spatial variations of substrate coverage. Since coverage inhomogeneities generate internal stresses, the coupling between coverage evolution and film elasticity fields is also considered, for film thickness below the critical thickness for misfit dislocation nucleation. It is shown that this coupling is destabilizing and favors nanostructure formation. It also favors square planforms which could compete, and even dominate over the haxagonal or stripe nanostructures induced by coverage dynamics alone.\\
\end{abstract}
\pacs{PACS numbers: 68.55.-a, 64.60.Cn, 81.15.-z, 05.65.+b}
\maketitle
\section{Introduction.}
The formation of self-assembled nanostructures in deposited layers on solid surfaces has become the subject of intense research activity, as a result of its fundamental and technological importance. Self-organized nanophases of different symmetries have been observed in binary epilayers (e.g. \cite{binary_1,binary_2}), but also in monoatomic layers (e.g. \cite{mono_1,mono_2}). Various atomistic computer simulation methods and continuous dynamical models have been developed to describe this phenomenon. Atomistic simulations are essentially based on adsorption of deposited atoms and their diffusion on the growing surface. They are performed with different methods, from molecular dynamics to Monte Carlo computations in continuum spaces, or discrete lattices. Attempts have also been made to bridge molecular dynamics and Monte Carlo methods, and succeeded in simulating small polycrystalline films \cite{hanchen} and the growth of thin films of larger dimensions \cite{cale}. These simulations provide essential information about the key factors which determine the properties of growing films, such as deposition kinetics, surface diffusion of atoms and defects, interatomic and surface potentials, etc. Since the size of the films described by these methods remains small due to computational limitations, continuous models remain of interest to describe mesoscopic scales (between the $\mu$m and the mm). Up to now, these models had limited predictive capability, because of a rough description of kinetic processes, such as atomic diffusion or deposition.  Nevertheless, such capability could be greatly enhanced in the framework of multiscale modeling. Effectively, the concept of Multiscale Materials Modeling has been recently introduced to bridge the gaps between atomistic and continuum methods, and to link them in a consistent way \cite{MMM,hanchenMRS}. The aim is to obtain a reliable description of materials behavior, from microscopic to macroscopic scales. This program should be realized by coupling models for different length scales. The results from smaller scales are fed to larger scales, with appropriate mesh redefinition, and the results from larger scales are being fed back to the smaller ones, in a back and forth process hopefully ending in quantitatively reliable solution. In the case of thin film growth, if information from each scale is transferred correctly to the other scales, one would expect to be able to make quantitative predictions on the evolution of film textures, surface topography, the effect of microstructures on local deposition rates, etc... .

Continuous models have already been proposed to describe the spontaneous ordering of nanostructures or self-assembled quantum dots in multicomponent epilayers on a substrate \cite{desai-97,desai-98,suo-00a,suo-00b}. They are based on an underlying instability of the alloy or solid solution which form the film. In this case, when the solid solution is unstable, below a critical temperature, it undergoes spinodal decomposition, which leads to phase coarsening. On the other hand, concentration-dependent surface stress or atomic deposition on the surface drive phase refining. As a result of the competition between these two effects, the phases sometimes select stable, nanometric, sizes. Furthermore, they may order into periodic patterns, such as alternating stripes or disks lattices. It has also been shown that, even in monocomponent films, the competition between atomic deposition and the underlying instability of an adsorbed atomic layer may generate nanoscale spatial patterns, already in the first deposited layers \cite{walgraef_02a,walgraef_02b}. These patterns correspond to regular spatial distributions of high and low coverage domains, which may induce corresponding distributions of grains with different orientations or symmetries, and serve as templates for the later stages of film textures evolution. Patterns with different symmetries may be selected, according to the relative values of experimental parameters such as deposition rate, substrate temperature and atomic mobility. Furthermore, for low deposition rates, patterns may change with time, according to the evolution of the mean surface coverage. In this case also, the selective formation of low or high coverage domains or islands is expected to generate internal stresses in the adsorbed layer, even in homoepitaxial growth. Since stress effects may affect nanostructure formation and stability and have thus to be incorporated in the dynamics. It is thus the purpose of the present paper to incorporate elasticity effects in our previous analysis of nanostructure formation during the early stages of thin film growth \cite{walgraef_02a,walgraef_02b}, and to show that stress effects alone may induce pattern forming instabilities, even in the absence of spinodal deposition like instability.

The paper is organized as follows. In section \ref{dynmodel}, a dynamical model describing the evolution of a deposited atomic layer on a substrate, and which takes into account adsorption, desorption, nonlinear diffusion, and coupling of atomic concentration or coverage with elasticity fields, is presented. The possible existence of a pattern forming instability in such a system is analyzed in section \ref{stability} and pattern selection is described in section \ref{patterns}. The relevance of the results to the interpretation of experimental observations or atomistic simulations is discussed in section \ref{check}. Finally, conclusions are drawn in section \ref{conclusions}.
\section{The Dynamics of a Deposited Layer on a Substrate.}\label{dynmodel}
In previous publications, a reaction-diffusion model, based on adsorption, desorption and nonlinear diffusion, was shown to be able to describe nanostructure formation in growing films on a substrate through evaporation or sputtering, for example\cite{walgraef_02a,walgraef_02b,walgraef03a}. However, the formation of domains, islands or nanostructures, with spatially varying atomic concentration, and which is usual in such films, is expected to generate internal stresses in the surface layers. Since these stresses may affect nanostructure formation and stability, they have to be incorporated in the layer dynamics, which should then be described, at the mesoscale, by coupled evolution equations for  atomic concentrations, or coverage, and layer deformation fields.

\begin{figure}
\includegraphics[width=4.5in]{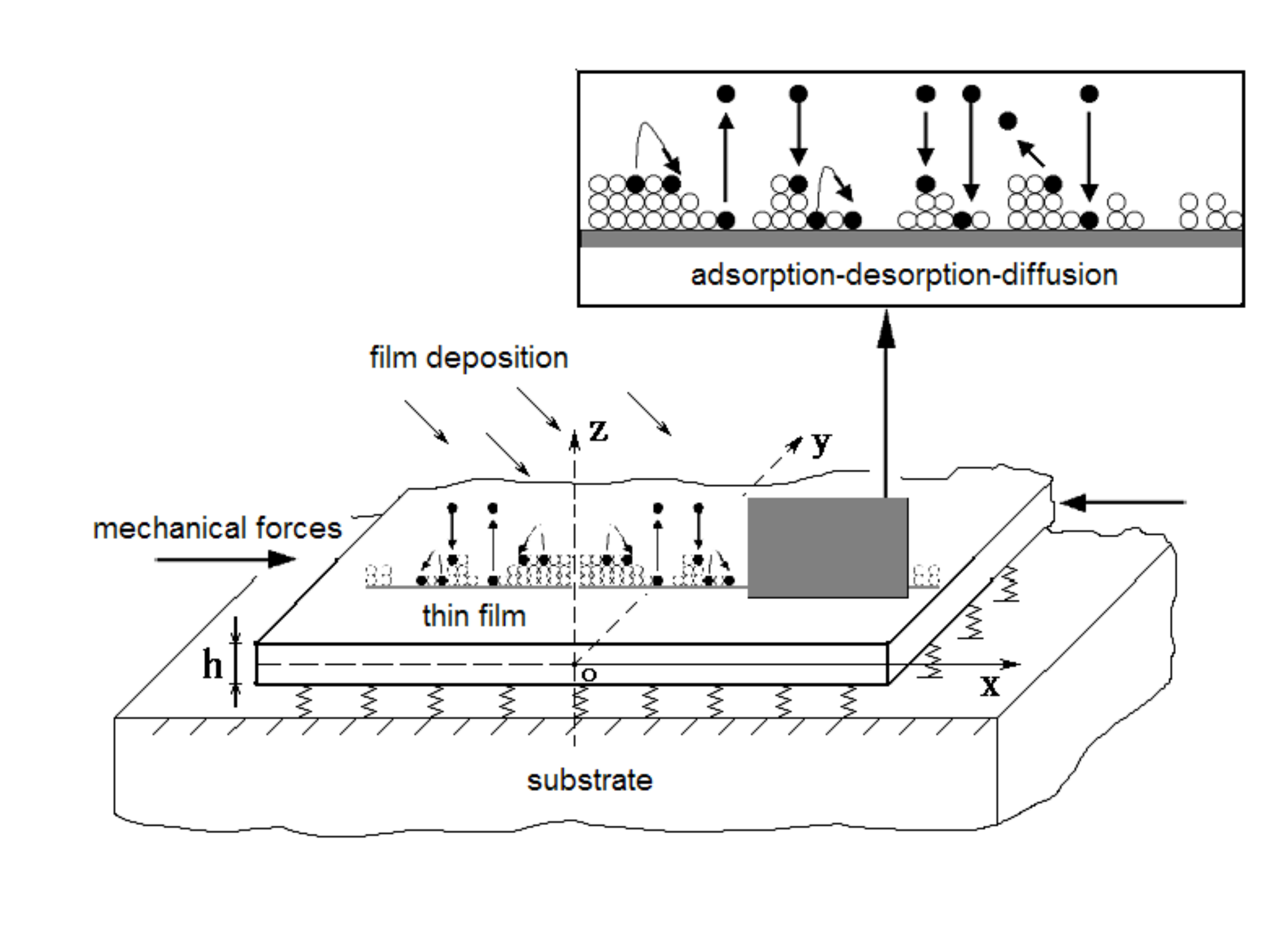}
\caption{Schematic representation of the basic mechanisms and geometrical factors involved in a minimal description of the deposition of a monoatomic film on a substrate.}
\label{setup}
 \end{figure}

The derivation of the deformation dynamics of a thin film on a substrate, in the presence of a spatially varying defect concentration field has been presented in \cite{lauzeral97}. The evolution equation for the film bending coordinate can be adapted to the present case, where vacant sites, deposited atoms and adatoms, islands, grains, etc. induce lattice dilatation (positive or negative) in the adsorbed layer. The corresponding elastic free energy may effectively be written as:
\begin{eqnarray}
{\cal F}_{el} &=& \int\int  \lbrack {Eh^3\over 24(1-\nu^2 )} \left( ( \triangle \xi)^2
+ 2(1-\nu )( \xi_{,xy}^2 - \xi_{,xx} \xi_{,yy} )\right) \nonumber\\
&+& {h\over 2} \epsilon_{\alpha\beta}\sigma_{\alpha\beta}\rbrack dS + \int\int\int^{h/2}_{-h/2} F_d dzdS
\end{eqnarray}
where $\xi$ is the bending coordinate of the film. $z$ is the film coordinate perpendicular to the substrate surface. For a horizontal film of thickness $h$, it varies from $-h/2$, which localizes the interface between film and substrate, and $h/2$, which localizes the film upper surface (cf. figure \ref{setup}). ($x,y$) are the planar coordinates parallel to the substrate surface. $\epsilon_{ij}$ and $\sigma_{ij}$ are the strain and strain tensors and $\xi_{,ij}$ stands for $\partial^2_{ij}\xi$ . $E$ is the film Young's modulus and $\nu$ its Poisson ratio.
$F_d$ is the energy stored in lattice dilatation inside the film. Such energy may be induced, for example, by point defect or variations in atomic coverage. For lattice dilatation induced by the presence of point defects, it is given, per unit volume, by:
\begin{equation}
F_d =  \sum_i c_i.(E^0_i + E_i^B + E^S_i)
\end{equation}
where $E^0_i$ is the self-energy per defect of type $i$, of concentration $c_i$, and $E^B_i$  and $E^S_i$ are the components of interaction energy due to bending and stretching respectively.  For vacancies, $E^0_v \simeq -\theta_v $, with $\theta_v \simeq 0.6 b^3K$, where $b$ is the Burger's vector and $K$ the bulk elasticity modulus. For adatoms, $E^0_a \simeq \theta_a $, with $\theta_a \simeq 0.3 b^3K$.

For bending\index{bending} only, one has $\partial_z U_z =\partial_z \xi =0$, and  $\vec \nabla . \vec U = -mz\triangle \xi$, with $m = {1-2\nu \over 1-\nu}$. Ignoring the contribution of the stretching energy in doing work on the strain field of defects, the total bending energy stored in lattice dilatation in the layer is then:
\begin{equation}
{\cal F}_d \simeq - \sum_i\int\int\int^{h/2}_{-h/2}A_i\,\theta_ic_i \, dzdS
\end{equation}
where $A_i = (1 + mz\triangle \xi  )$.

The corresponding free energy variations are thus
\begin{eqnarray}
\delta {\cal F}_d &\simeq & -\sum_i \theta_i \int\int\int^{h/2}_{-h/2}\biggl( ({\partial c_i\over \partial z} + zm\triangle c_i  )\delta\xi + A_i\delta c_i \biggr) dzdS\nonumber\\
&=&-\sum_i \theta_i \int\int \Biggl[ \biggl( c_i(+)-c_i(-) + m \triangle I_i(c_i)\biggr)\delta\xi + \int^{h/2}_{-h/2}dzA_i \delta c_i\Biggr] dS
\end{eqnarray}
where $c_i{\pm} = c_i(x,y,\pm \frac h2)$ and $I_i(c_i)= \int^{h/2}_{-h/2} z c_i dz$. Vacancy concentration, $c_v$, may be related to local atomic coverage, $c$, by $c_v = 1 - c$, and since atomic coverage is the relevant variable for thin film growth, the kinetic equation for film bending  becomes \cite{lauzeral97}:
\begin{equation}\label{bendingdyn}
\partial^2_t \xi + {Eh^2\over 12\rho (1-\nu^2)}\triangle^2 \xi
-{1\over \rho} (\sigma_{\alpha \beta}\xi_{,\alpha})_{,\beta} =
{\theta \over b^3\rho h}[c_+-c_- + m \triangle I(c) ] + {F(\xi )\over \rho h}
\end{equation}
where $c_+$ and $c_-$ are the atomic concentrations at the layer top and bottom surfaces, respectively. $F(\xi )$ is the adhesive force, between film and substrate, per surface unit. The force of adhesion between the bottom surface of the film and the top layer may be represented by the universal bonding curve, usually invoked in process zone fracture models \cite{needleman,xu,siegmund}. However, since the film instability threshold is governed by the small displacement part of the $F(\xi )$ function, it may be approximated by  $F(\xi )= - \kappa\xi$, where $\kappa$ is the adhesive bond stiffness constant.

Since the film is thin, the in-plane first Piola stress tensor variation will be ignored. This will allow us to re-write the third term in equation (\ref{bendingdyn}) as:
\begin{equation}
{1\over \rho} \sigma_{\alpha \beta}\xi_{,\alpha})_{,\beta} =
{1\over \rho} \sigma_{\alpha \beta}\xi_{,\alpha\beta}
\end{equation}
where the in-plane stress tensor $\bar\sigma$ is given by
\begin{eqnarray}
\sigma_{xx} &=& \sigma_m + {E\over 2(1-\nu^2)} [\xi^2_{,x} + \nu \xi^2_{,y}
+(1+\nu )\alpha \Delta T + z N_{xx}]\nonumber\\
\sigma_{yy} &=& \sigma_m + {E\over 2(1-\nu^2)} [\xi^2_{,y} + \nu \xi^2_{,x}
+(1+\nu )\alpha \Delta T + z N_{yy}]\nonumber\\
\sigma_{xy} &=& {E\over 1+\nu} [\xi_{,x}\xi_{,y}+(1+\nu )\alpha \Delta T + z N_{xy}]
\end{eqnarray}
where $\sigma_m = {E\over 2(1-\nu )}\epsilon_m$, where $\epsilon_m$ is the misfit strain induced by an isotropic substrate on the deposited layer, and
\begin{equation}
N_{\alpha\beta} = (1-\nu ) \xi_{,\alpha\beta} + \nu
\delta_{\alpha\beta}
\xi_{,\alpha\alpha}
\end{equation}
In these relations, the stretching\index{stretching}, thermal and bending strains are included.The thermal expansion coefficient is $\alpha$ and $\Delta T$ is the average (across the thickness) temperature rise in the film, and may be neglected in the present case, and we finally obtain:
\begin{equation}\label{bendingkineq}
\partial^2_t \xi + {Eh^2\over 12\rho (1-\nu^2)}\triangle^2 \xi -{1\over 2\rho } \sigma_{\alpha \beta}\xi_{,\alpha\beta} = {\theta \over b^3\rho h}[c_+ - c_- + m \triangle I(c) ]-{\kappa\over\rho h}\xi
\end{equation}
From equation (\ref{bendingkineq}), one sees that the film bending evolution is coupled to atomic coverage at the top and bottom surfaces. In a previous publication, the growth of a deposited film on a substrate has been described by a reaction-diffusion model, where atoms are adsorbed and desorbed, but also diffuse in the film \cite{walgraef03a}. Let us recall the basic elements of this model, which are of interest here. Multilayer films are considered, where successive layers are deposited on top of each other. In a specific layer, atoms are adsorbed on unoccupied sites, but on occupied sites of the previous underlying layer, while atoms are desorbed from occupied lattice sites. Both mechanisms may occur provided the corresponding sites of the upper layers are vacant. Hence, for a film made of $n$ layers, where individual layers are labeled by the integer $i$ going from 1 to $n$, the dynamics of bulk layers is accordingly given by ($1\le i < n$):
 \begin{eqnarray}\label{bulklayerNLD}
\partial_t c_i &=& [\alpha c_{i-1}(1-c_i )- \beta c_i](1-c_{i+1} )  - \vec\nabla\vec J_i + D_z(c_{i+1}+c_{i-1}-2c_{i})\nonumber\\
\vec J_i &=& -D_h \vec\nabla \lbrack  - \frac{\epsilon_0}{k_BT} c_i  + \ln {c_i \over 1-c_i
} - \frac{\xi_0^2}{k_BT}\nabla^2c_i\rbrack
\end{eqnarray}
where $D_z$ is the vertical diffusion coefficient, which is much smaller than the lateral diffusion coefficient ($D_z<<D_h$), and $c_{0}=1$ and $c_{n+1}=0$. The $1-c_{i+1}$ term reflects the absence of adsorption and desorption inside the film. It is negligeable in the bulk and close to one near the surface. Assuming local thermodynamical equilibrium, $J_i$ is the current induced by the free energy of an adsorbed atomic layer \cite{walgraef_02a,walgraef_02b}. On the other hand, the evolution of the surface layer is considered to be given by the following kinetic equation:
\begin{eqnarray}\label{surfacelayerNLD}
\partial_t c_n &=& \alpha c_{n-1}(1-c_n ) - \beta c_n - \vec\nabla\vec J_n + D_z(c_{n-1}-c_{n})\nonumber\\
\vec J_n &=& -D_h \vec\nabla \lbrack  - \frac{\epsilon_0}{k_BT} c_n  + \ln {c_n \over 1-c_n
} - \frac{\xi_0^2}{k_BT}\nabla^2c_n\rbrack
\end{eqnarray}
where $c_n$ is the corresponding atomic coverage.

In the present case, we are considering multilayer films which may be described by elasticity free energy, which implies that $n>>1$, or at least $n\ge 1000$. Hence, one may consider the continuous limit of the system (\ref{surfacelayerNLD},\ref{bulklayerNLD}), and, on defining $c_i(\vec r,t)= c(z,\vec r,t)\vert_{z=ia}$, where $a$ is the lattice constant, equations (\ref{bulklayerNLD},\ref{surfacelayerNLD}) may be cast into:
\begin{eqnarray}\label{pdebulkdyn}
\partial_t c &=& [\alpha (1-c )- \beta ]c - \alpha (1-c)a\partial_z c + D^*_z\partial^2_z c  \nonumber\\
&+& D_h \nabla^2 \lbrack  - \frac{\epsilon_0}{k_BT} c  + \ln {c \over 1-c} - \frac{\xi_0^2}{k_BT}\nabla^2c\rbrack
\end{eqnarray}
where $D^*_z = D_z + \frac{a^2}{2}\alpha (1-c)$ and $c(z\to\infty ,\vec r,t)=0$.

\begin{figure}
\includegraphics[width=9cm]{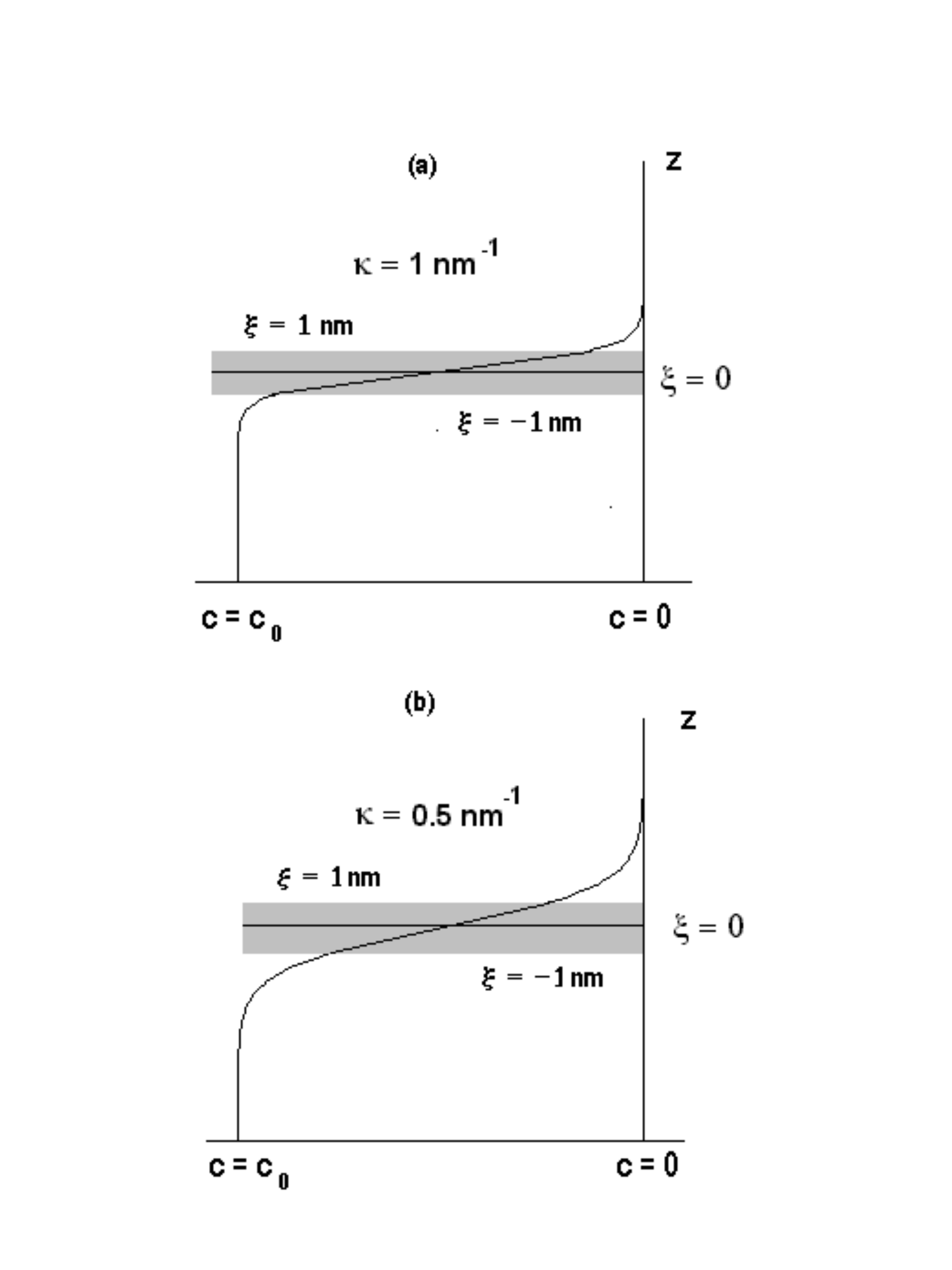}
\caption{Horizontal concentration profile corresponding to the pulled front solution of eq. (\ref{vertgrowth}) for $D^*_z$ = 0.5 $\times$ 10$^{-8}$ cm$^{2}$s$^{-1}$, $a = 4 \AA$, (a) $\alpha$ = 200 $\mu$m s$^{-1}$, (b) $\alpha$ = 50 $\mu$m s$^{-1}$, and $\beta < 0.1 \alpha $, showing that the surface layer is sharper for larger adsorption rate.}
\label{front}
 \end{figure}

The evolution of a laterally uniform growing film, as deduced from eq. (\ref{pdebulkdyn}), is given by:
\begin{equation}\label{vertgrowth}
\partial_t c_0(z,t) = [\alpha (1-c_0(z,t) )- \beta ]c_0(z,t) - \alpha (1-c_0(z,t))a\partial_z c_0(z,t) + D^*_z\partial^2_z c_0(z,t)
\end{equation}
where $c_0(z,t)$ is the laterally uniform coverage, with $c_0(0,t\to\infty )\to \frac{\alpha - \beta}{\alpha}$ and $c_0(z\to \infty ,t)=0$. Hence, $c_0(z,t)$ tends to a moving front solution connecting the stable steady state $c_0 = \frac{\alpha - \beta}{\alpha}$ at $z=0$ to the unstable steady state $c_0=0$ at $z\to \infty$. This is a so-called "pulled" front. It is governed by its leading edge dynamics \cite{wim1}, which is given by
\begin{equation}\label{leadingedgevertgrowth}
\partial_t c_0(z,t) = \alpha c_0(z,t)- a\alpha \partial_z c_0(z,t) + D^*_z\partial^2_z c_0(z,t)
\end{equation}
where $c_0(z,t)= c_0(\zeta )\simeq \exp k\zeta$ where $\zeta = z - vt$. Its propagation velocity and decay are given by the marginal stability criterion \cite{wim1} :
\begin{eqnarray}\label{dispersionvertgrowth}
 \alpha + (v_0 - a\alpha )k_0 + D^*_zk_0^2 &=& 0\nonumber\\
(v_0 - a\alpha ) + 2D^*_zk_0 &=& 0
\end{eqnarray}
or
\begin{equation}
k_0 = - \sqrt{{\alpha \over D^*_z}}\,{\rm and}\, v_0 = a\alpha + 2\sqrt{\alpha D^*_z}
\end{equation}
As a result, $c_0(z,t)=c_0(\zeta )\simeq \exp k_0\zeta$, for $\zeta >0$, where $\zeta = z - v_0t$ with  and while $c_0(\zeta )\simeq \frac{\alpha - \beta}{\alpha}$, for $\zeta < 0$. On the other hand, the upper surface of the growing film, corresponds to the pulled front $c_0(\zeta )$ at $\zeta\simeq 0$ (cf. figure \ref{front}).

To allow for lateral spatial variations, which have not been considered up to now, let us define
$c(z,\vec r,t) = c_0(\zeta )\phi (\zeta , \vec r, t)$. The evolution of $\phi$, derived from (\ref{pdebulkdyn}), is then given by
\begin{equation}\label{pdelateraldyn}
\partial_t \phi  =  \alpha c_0(\zeta )(1-\phi )\phi + D^*_z\partial^2_{\zeta } \phi  + D_h \nabla^2 \lbrack   \frac{1}{c_0(\zeta )}\ln\frac{\phi }{ 1-c_0(\zeta )\phi }- \frac{\epsilon_0}{k_BT} \phi  - \frac{\xi_0^2}{k_BT}\nabla^2\phi \rbrack
\end{equation}

\begin{equation}
c_- = c_0(\zeta < 0)\phi (0,\vec r,t)=\phi_-\,\,,\,\,c_+ = c_0(0)\phi (0,\vec r,t)=\frac{\alpha - \beta}{\alpha}\phi_+
\end{equation}
For small desorption rates (which is the case usually considered in numerical simulations where desorption is neglected), $c_0(\zeta < 0)\simeq 1 \to \phi_- =1$
Neglecting vertical diffusion and interactions, and incorporating the effect of the interaction free energy, ${\cal F}_d$, into the current, the complete dynamical model finally becomes:
\begin{eqnarray}\label{stressmodel}
\partial^2_t \xi &=&- {Eh^2\over 12\rho (1-\nu^2)}\triangle^2 \xi +{1\over 2\rho } \sigma_{\alpha \beta}\xi_{,\alpha\beta} -{\kappa\over\rho h}\xi + {\theta \over b^3\rho h}c_0(0)(\phi_+ - 1 )\nonumber\\
\partial_t \phi_+ &=&  \frac{1}{\tau}(1-\phi_+ )\phi_+  \nonumber\\
 &+& D_h \nabla^2 \lbrack   \frac{1}{c_0(0 )}\ln\frac{\phi_+ }{ 1-c_0(0 )\phi_+ }- \frac{\epsilon_0}{k_BT} \phi_+  - \frac{\xi_0^2}{k_BT}\nabla^2\phi_+ +\frac{mh\theta}{2k_BT}\nabla^2 \xi  \rbrack
\end{eqnarray}
where $c_0 = \frac{\alpha -\beta}{\alpha}$ and $\tau^{-1} = \alpha - \beta$. Fick diffusion and local thermodynamic equilibrium have been considered in the derivation of (\ref{stressmodel}). For the sake of simplicity, surface diffusion has been considered as isotropic, which is the case for most of the experimental systems we will investigate. Kinetic equations for hopping types of atomic motion will be analyzed in a separate publication. To facilitate further analysis, equation (\ref{stressmodel}) may be expressed in scaled variables, on defining
\begin{eqnarray}\label{scaled}
\partial_T = \tau \partial_t \, &,& \, \bar\triangle = \frac{4T_cD_h\tau}{T}\triangle \,\, , \,\, l^2 = \frac{n_0a^2_0T}{4T_cD_h\tau} \nonumber\\
\bar\xi = \frac{mh\theta T}{32k_BT_c^2D_h\tau}\xi \,\, &,& \,\,
\Lambda = {Eh^2T^2\over 192\rho (1-\nu^2)D_h^2T_c^2} \,\, , \,\,  \bar\epsilon_{m,ij} = \frac{24T_cD_h\tau}{Th^2}\epsilon_{m,ij}     \nonumber\\
  Q= {\kappa\tau^2\over\rho h\Lambda}        \,\, &,& \,\, u = {6m(1-\nu^2 )\theta^2D_h\tau c_0(0)\over Eh^2b^3k_BT_c}
\end{eqnarray}
where $n_0$ is the lattice coordination number and $a_0$ the lattice constant. This yields
\begin{eqnarray}\label{scaledstressmodel}
\frac{1}{\Lambda}\partial^2_T \bar\xi &=&- \bar\triangle^2 \bar\xi + \bar\epsilon_{m,ij}\bar\nabla_i\bar\nabla_j\bar\xi -Q\bar\xi + u(\phi_+ - 1 )\nonumber\\
\partial_T \phi_+ &=& (1-\phi_+ )\phi_+  \nonumber\\
 &+& \bar\triangle \lbrack  \mu(1 - c_0(0 ))\ln\frac{\phi_+ }{ 1-c_0(0 )\phi_+ }- \phi_+  - l^2\bar\triangle\phi_+ +\bar\triangle \bar\xi  \rbrack
\end{eqnarray}
\section{Strain induced instability of a growing layer.}\label{stability}
The system (\ref{stressmodel}) admits a uniform steady state, $\phi_+ = 1$, $\xi = 0$, which corresponds to a uniform and undeformed deposited layer. Its stability versus inhomogeneous coverage and deformation perturbations is given by the linear evolution of small perturbations $\varphi = \phi_+ -  1$ and $\xi$, which may be written, in Fourier transform, as:
\begin{eqnarray}
\partial_T\varphi_q &=& -\lbrack 1 + (\mu -1)q^2  + q^4l^2\rbrack \varphi_q + q^4 \bar\xi_q\nonumber\\
\frac{1}{\Lambda}\partial^2_T\bar\xi_q &=& - \lbrack q^4   + \bar\epsilon_m q^2 + Q \rbrack\bar\xi_q + u \varphi_q
\end{eqnarray}
where the misfit strain has been considered as isotropic.
The eigenvalues of the corresponding linear evolution matrix are the solutions of the characteristic equation
\begin{equation}\label{roots}
\lbrack\omega + 1 + (\mu -1)q^2  + q^4l^2 \rbrack . \lbrack\omega^2 + q^4   + \bar\epsilon_m q^2 + Q \rbrack = q^4u
\end{equation}
In the absence of coupling between atomic coverage and deformation fields, instability may be due to the adsorbed layer instability, as studied in \cite{walgraef_02a,walgraef_02b,walgraef03a}, or to compressive misfit strains such that
\begin{equation}
\vert\epsilon_m\vert > \sqrt{\frac{16\kappa}{3E}.\frac{1-\nu}{1+\nu}. h}
\end{equation}
In this latter case, spatial modulations of unscaled wavelength
\begin{equation}
\lambda_0 = 2\pi\sqrt{\frac{2h^2}{3(1+\nu)\vert\epsilon_m\vert}}
\end{equation}
have maximum growth rate and are linearly selected. For deposited films of thickness around the micrometer, with elasticity modulus around $E\simeq 70 GPa$, such as for $Al$, and adherence stiffness constant $\kappa \simeq 25 MPa/m$, instability occurs for compressive misfit strains such that $\vert\epsilon_m\vert \ge 10^{-4}$. The corresponding wavelength is of the order of 100$h$ and decreases for increasing misfit strain intensity. Note that this analysis is only valid for elastic film deformation, which occurs in films thinner than the critical thickness above which misfit dislocations develop.

Since the critical wavelength, at instability threshold, is given by
\begin{equation}
\lambda_c = 2\pi(\frac{Eh^3}{12\kappa (1-\nu^2)})^{1/4}\, ,
\end{equation}
it corresponds to a wavelength which grows with the layer thickness as $h^{3/4}$, for constant elasticity modulus and adherence forces. As discussed in \cite{lauzeral97} elastic nonlinearities are of the Proctor-Sivashinsky type, and the nonlinearly selected deformation patterns correspond to square lattices with wavelength in the micrometer range for misfit strains of the order of 0.1\%. In this case, the critical thickness is in the range 0.01\%.

On the other hand, the coupling between atomic coverage variations and deformation fields has a destabilizing effect on uniform coverage. It effectively lowers the decay rate of spatial modes in films which are otherwise stable with respect to elastic deformations and spinodal decomposition.
For temperatures above critical, ($T \ge 4T_cc_0(1-c_0)$ or $\mu > 1$), the film is stable versus spinodal decomposition, and $\sigma_q$ may be adiabatically eliminated. Instability then occurs for
\begin{equation}\label{instroot}
 Q  + \bar\epsilon_m q^2 + q^4  < q^4\frac{u}{1 + (\mu - 1) q^2   + l^2q^4 }
\end{equation}
  This effect may induce instability, even for $\epsilon_m \ge 0$. A necessary conditions for instability is, in this case, in unscaled units:
\begin{equation}\label{thickcond}
\frac{\xi_0^2b^3 \kappa }{m\theta^2} < h <\sqrt{{12 (1-\nu^2)m\theta^2D\tau \over Eb^3k_BT}}
\end{equation}
which confirms the qualitative observation that the destabilizing effect of the coupling vanishes at small and large layer thickness. For realistic values of the physical parameters, such as in the deposition of $Al$ on $TiN$, with a deposition rate of the order of $\mu$m/min at $300^0 $K, $D \simeq 2.10^{-7}$ cm$^2$s$^{-1}$ and $a_0 \simeq 4\AA$, condition (\ref{thickcond}) yields $a_0\le h < 1.2 \, \mu$m (note that, for monolayers ($h\to a_0$), the film elasticity theory used so far is not valid any more, due to size effects).

Let us consider more precisely the destabilization effect of the deformation-deposition coupling close to the adsorbed layer instability, in the case of homo- and heteroepitaxy.
\subsection{Strain induced instability of epitaxial layers}
In the case of homoepitaxy, $\epsilon_m = 0$, and uniform steady states are unstable, close the adsorbed layer instability temperature ($T\simeq 4T_cc_0(1-c_0)$ or $\mu \simeq 1$), for
\begin{equation}\label{homoroot}
 Q + q^4 < q^4\frac{u}{1 + q^4l^2 }
\end{equation}
Uniform layers are thus unstable for $\sqrt{u} > 1 + \sqrt{Q}$ and the maximum growth rate corresponds to modulations of wavenumber $q_0= \frac{\sqrt u - 1}{l^2}$. This condition implies that instability may only occur for deposited layers of thickness $h$, such that
\begin{equation}
L_0 < 3({h\over h_0})^{1/2} - ({h\over h_0})^{3/2} < 2
\end{equation}
where $h_0^2= \frac{4 (1-\nu^2)m\theta^2D\tau}{3k_BTEb^3}$ and $L_0 =({27\over 2}\kappa\xi_0^2)^{1/2}({Ek_BT\over 3(1-\nu^2)D\tau})^{1/4}({b^3\over m\theta^2})^{3/4} $. Hence,  layers in the range
\begin{equation}
h_- < h < h_+
\end{equation}
where $h_{\pm}= 2h_0\cos \frac{\pi\mp\phi}{3}$, with $\phi = \arctan \sqrt{\frac{4-L_0^2}{L_0^2}}$, are unstable.
 The critical wavelength is defined by
\begin{equation}
\lambda_c = 2\pi(\frac{\xi_0^2D\tau Eh^3}{12(1-\nu^2)\kappa k_BT})^{1/8}
\end{equation}
and scales as $h^{3/8}$.
Instability may occur for $L_0<2$ only, which sets a limit for adhesion strength, above which films are stable. When  $L_0\le 2$, films of thickness $h\simeq h_0$ may be unstable, while for small $L_0$ ($L_0\to 0$), films of thickness in the range $a_0<h<\sqrt 3 h_0$ may be unstable.
 For $Al$ deposited on $TiN$, in the conditions described previously, $h_0\simeq 1 \mu m$ and $\lambda_c\simeq 25 \, \mu$m.
\subsection{Strain induced instability of heteroepitaxial layers}
In the case of heteroepitaxial deposition, $\epsilon_m \ne 0$, and uniform layers are unstable, close to the adsorbed layer instability temperature ($T\simeq 4T_cc_0(1-c_0)$ or $\mu\simeq 1$), for
\begin{equation}\label{heteroroot}
 Q + q^2 \bar\epsilon_m + q^4(1 - \frac{u}{1 + q^4 l^2})< 0
  \end{equation}
For films such that $h > \sqrt 3 h_0$, instability occurs for
\begin{equation}\label{heteroroot2}
\epsilon_m  \le - \sqrt{\frac{16\kappa (1-\nu )h}{3E(1+\nu )}}.\sqrt{1-{3h_0^2\over h^2}}
  \end{equation}
For films such that $h < \sqrt 3 h_0$,
\begin{equation}\label{heteroroot3}
\epsilon_m  \le  \sqrt{\frac{16\kappa (1-\nu )h}{3E(1+\nu )}}.\sqrt{({3h_0^2\over h^2}-1)}
  \end{equation}
Expressions (\ref{heteroroot2}) and (\ref{heteroroot3}) show that the deformation-coverage coupling increases the instability range and may even induce instability for positive misfit strains.
 For $Al$ deposited on $TiN$, in the conditions described earlier, this corresponds to $\epsilon_m  \le  10^{-4}$.
\section{Elasticity effects on the self-organization of a deposited layer.}\label{patterns}
\begin{figure}
\includegraphics[width=4in]{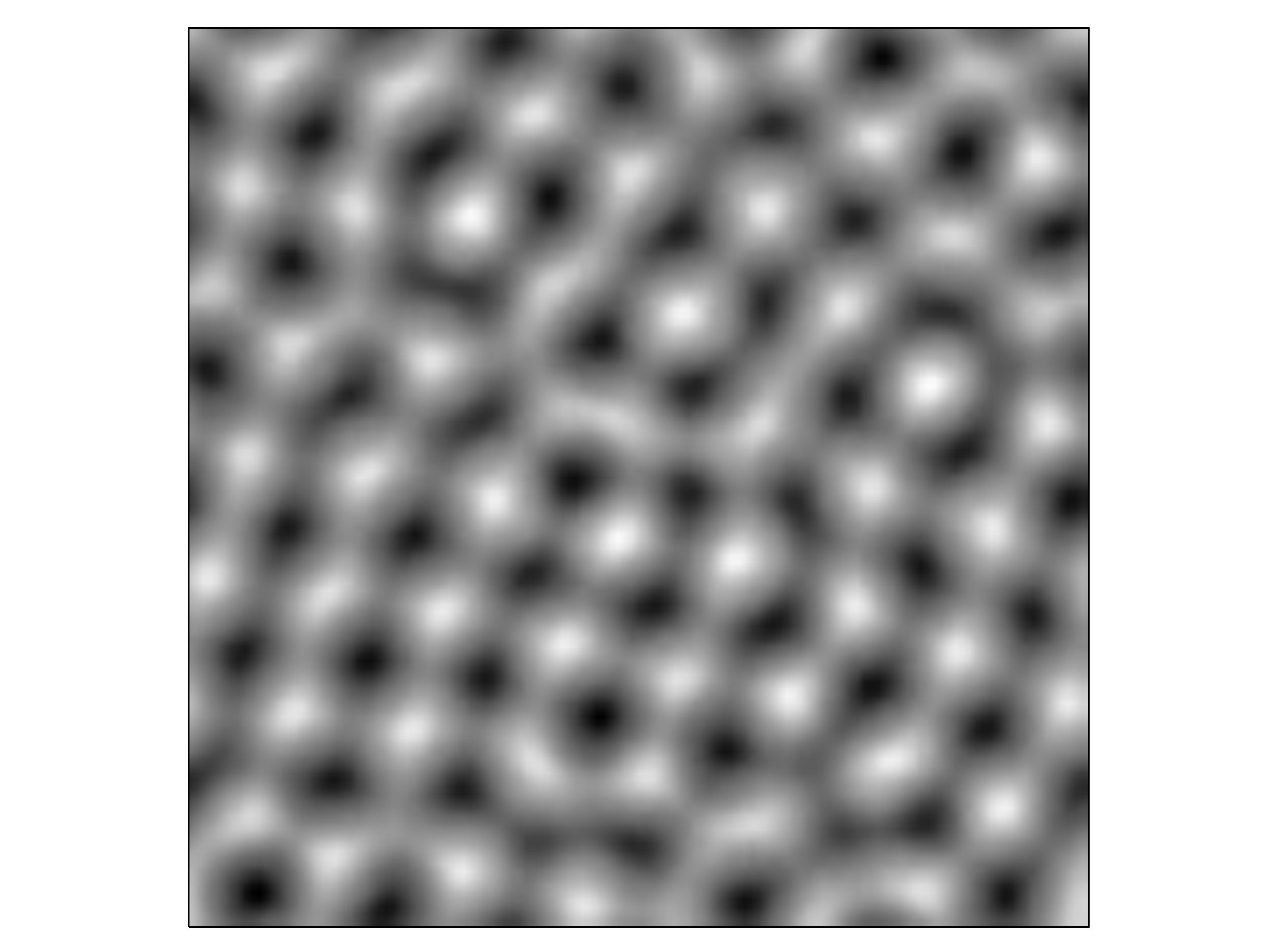}
\caption{Square-like patterns obtained in the numerical analysis of the order parameterlike equation (\ref{orderparameter}) for dominant elastic nonlinearities.
}
\label{squares}
\end{figure}

Let us consider now the effect of deposition-deformation coupling in temperature ranges where uniform deposited layers are unstable versus spinodal decomposition, but stable versus deformation, such as for negligeable misfit strains, where there is no intrinsic elastic instability, and where the critical thickness is of the order of the $\mu$m . Since $\Lambda >>1$ in realistic  experimental conditions (for 1 $\mu$m $Al$ films deposited at $300^0 K$, $\Lambda\simeq 5.10^{12}$), $\bar\xi_q$ may be adiabatically eliminated, and the relevant root for instability becomes:
\begin{equation}\label{instroot2homo}
\omega = - 1 - q^2 (\mu - 1) - q^4 \lbrack \, l^2 - \frac{u}{ Q + q^2 \bar\epsilon_m + q^4 }\rbrack
\end{equation}
It is easy to see that the deposition-deformation coupling shifts the instability temperature towards higher values, increasing the instability domain of uniform coverage. Analytical expressions may be obtained for this shifts in the strong and weak film-substrate adherence limits. However, the most significant elasticity effect on nanostructure formation lies in the nonlinear terms. On the one hand, the adiabatic elimination of spatial deformation modes yields the following expression, where $\xi_q$ is expressed as a series expansion in powers of spatial coverage variations $\varphi_q$ ($\vert \vec q\vert \simeq q_c$):
\begin{equation}\label{nolistressmodel}
\bar\xi_{\vec q} =    \gamma (q_c)\varphi_{\vec q}
+\gamma(q_c)^3\int_c d\vec q_1\int_c d\vec q_2 \, \bar u(\vec q , \vec q_1 , \vec q_2)\, \varphi_{\vec q - \vec q_1 -\vec q_2}\varphi_{\vec q_1}\varphi_{\vec q_2} + \dots
\end{equation}
where $\bar u(\vec q , \vec q_1 , \vec q_2) = (\frac{32k_BT_c^2D_h\tau}{mh\theta T})^2\lbrack {\nu\over  1+\nu } (\vec q - \vec q_1 -\vec q_2)^2(\vec q_1 . \vec q_2)^2 + {1-\nu \over  1+\nu }((\vec q - \vec q_1 -\vec q_2).\vec q_1)((\vec q - \vec q_1 -\vec q_2).\vec q_2)\rbrack$, and $\gamma (q)= \frac{u}{ q^4   + q^2 \bar\epsilon_m  + Q } $. The subscript $c$ indicates that the integral runs over the critical shell. The resulting order parameterlike equation reads
\begin{eqnarray}\label{orderparameter}
\partial_t\varphi_q &=& -\lbrack 1 + q^2 (\mu - 1) + q^4 (l^2 - \gamma (q))\rbrack \varphi_q  \nonumber\\ &-&\int_c d\vec q_1 \, v(\vec q )\,\varphi_{\vec q - \vec q_1 }\varphi_{\vec q_1}-
\int_c d\vec q_1\int_c d\vec q_2 \, w(\vec q , \vec q_1 , \vec q_2)\, \varphi_{\vec q - \vec q_1 -\vec q_2}\varphi_{\vec q_1}\varphi_{\vec q_2} + \dots
\end{eqnarray}
where $v(\vec q ) = - \frac{q^2\mu (1-2c_0(0))}{2(1-c_0(0))}$ and  $w(\{\vec q_i\}) =  \frac{q^2 \mu (1-3c_0(0)+3c_0(0)^2)}{3(1-c_0(0))^2}- \gamma(q_c)^3\bar u(\{\vec q_i\})$
In this equation, quadratic nonlinearities come from the deposition-diffusion part of the dynamics, while cubic nonlinearities contain a deposition-diffusion part and another part coming from elasticity effects, which is of the Proctor-Sivashinsky type.  The kinetic coefficient of the deposition-diffusion part of the cubic nonlinearity is proportional to $T/T_c$ while the kinetic coefficient of the elasticity part is proportional to $(T_c/T)^2$. As a result, the importance of  elasticity effects increases for decreasing temperatures. Hence, this dynamics may give rise to different pattern formation scenarios \cite{walgraef_96,lauzeral97}.
\begin{itemize}
\item Close to critical coverage ($c_0(0) \simeq 0.5$), quadratic nonlinearities are negligeable. For small elasticity contributions to cubic nonlinearities, selected patterns should correspond to bands, while for dominant elastic contributions, they should correspond to squares.
\item Far from critical coverage ($c_0(0) \simeq 0, 1$), quadratic nonlinearities are important and induce, slightly below instability temperature, hexagonal patterns corresponding to high coverage islands in a low coverage background ($c_0(0) \simeq 0$) or to low coverage islands in a high coverage background ($c_0(0) \simeq 1$). On lowering temperature, these patterns should transform into bands or squares (cf. fig. \ref{squares}), according to the importance of elastic contributions to cubic nonlinearities.
\end{itemize}

\section{Discussion.}\label{check}
In the preceeding sections, it has been shown how the evolution of a deposited film, modeled by the dynamical system (\ref{stressmodel}), may generate spatial patterns induced by elastic instability or by adsorbed layer instability. In the first case, the patterns should correspond, for isotropic elasticity, to square lattices with wavelengths in the micrometer range. In the second case wavelengths are expected to be in the nanometer range, but pattern selection is more intricate.

In fact, the system (\ref{stressmodel}) may be reduced, close to instability, to the order parameter equation (\ref{orderparameter}), which allows a qualitative analysis of pattern selection and stability. It is now well known that pattern symmetries are selected by nonlinear couplings between unstable modes. In equation (\ref{orderparameter}), nonlinearities are generated on the one hand by nonlinear diffusion of the Cahn-Hilliard type, and, on the other hand, by nonlinear elasticity. The first terms favor hexagonal or stripe structures, while the second ones favor square planforms, which should thus be selected if elastic nonlinearities are dominant.

Since all kinetic coefficients are complicated functions of materials parameters, deposition rates and temperature, a quantitative evaluation of the relative importance of elastic nonlinearities and elasticity effects can only be performed for specific experimental situations, which is out of the scope of the present paper. Nevertheless, some aspects of the discussion may be illustrated for $Al$ films deposited on $TiN$ for example. In particular, a dimensional analysis of the system shows (\ref{stressmodel}) that elastic effects become relevant for

\begin{equation}
{3(1-\nu )^2\theta^2\lambda^2\over 2\pi^2k_BTEb^3h^2 }> 1
\end{equation}
Elastic nonlinearities should thus become important for nanostructures of wavelength larger than about 20 nm, in $0.1 \mu$m thick homoepitaxial $Al$ films at room temperature. This "characteristic" wavelength is proportional to the film thickness and increases with the square root of the temperature. For wavelengths in the nanometer range, elasticity effects increase for decreasing temperature. Hence, for fixed materials parameters and deposition conditions, hexagonal or square nanostructures could thus be expected, according to the temperature of the growing film.

\section{Conclusions.}\label{conclusions}

In this paper, the evolution of a growing monoatomic layer, deposited on a substrate, has been described by a dynamical model of the reaction-diffusion type. This dynamics combines reaction terms (adsorption and desorption) and nonlinear diffusion, and, close to the critical point of the order-disorder transition of the adsorbed layer, it corresponds to modified Cahn-Hilliard equations. The result is that uniform layers may become spatially unstable at sufficiently low temperatures, and sufficiently high atomic mobility. In these conditions, it develops nanoscale spatial patterns, corresponding to regular distributions of high and low coverage domains in monoatomic layers, or domains of different species for polyatomic layers. Since on the one hand coverage or concentration variations generate internal stresses, and on the other hand coupling with the substrate generates misfit strains, coverage dynamics has been coupled with film elasticity fields. Their dynamical behavior has been described by the evolution of the film bending coordinate in the presence of coverage variations in the film surface.

It has been shown that the coupling between surface coverage variations and deformation fields is destabilizing and may induce or favor the formation of nanostructures. Several limits have been considered. Above the critical temperature of adsorbed layer instability, nanostructure formation may be induced either by misfit strains or by the coupling with coverage inhomogeneities. Preferred structures should correspond in this case to square lattices in isotropic systems. In the adsorbed layer instability domain, the coupling between coverage inhomogeneities and film deformation is twofold. On the one hand, it enlarges the linear instability domain. On the other hand, it adds new contributions to the nonlinear terms of the order parameterlike equation. These contributions may modify the selected nanostructures since they favor square lattices. Hence, when coverage nonlinearities dominate, selected nanostructures should correspond to hexagons, or wrinkles when the mean coverage is close to critical. On the contrary, when deformation nonlinearities dominate square planforms should be selected. The preferred wavelength is expected to decrease proportionally to the inverse of the square root of the amplitude of the misfit strain. For 1\%, this wavelength should be around 50 h, where h is the film thickness. For ultrathin films, our results are reminiscent of Stranski-Krastanov growth, although they are strongly dependent on dynamical aspects. Note that films of thickness below the critical thickness for nucleation of misfit dislocations have been considered throughout this paper. Necertheless, for increasing film thickness, the threshold for the formation of such dislocations may be reached. In this case, it would be interesting to know if preexisting deformation patterns, such as the ones discussed here, could affect the spatial distribution of these dislocations, and act as templates for the self-assembly of nanostructures or quantum dots.

Instability thresholds, critical temperatures, wavelengths, etc., may be expressed as functions of experimental parameters such as deposition or adsorption rates, substrate temperature,  surface diffusion coefficients, and mechanical parameters such as the film Young´s modulus, Poisson ratio, and misfit strains. Hence, quantitative results may be obtained for specific systems and could allow for comparisons with experimental data. These aspects will be addressed in subsequent publications as well as the problem of elasticity effects on texture formation. It is also worth noting that, since the dynamics analyzed here does not derive from a potential, patterns wavelengths and symmetries are selected by dynamical processes, and not by variational principles. From an experimental point of view, this allows more flexibility in the designing and processing of self-assembled nanostructures.

\begin{acknowledgments}
Fruitful discussions with Profs. N.M. Ghoniem, H. Huang and E.C. Aifantis are gratefully acknowledged.
\end{acknowledgments}
\bibliography{ThinFilms}

\end{document}